\documentclass[sigconf]{acmart}

\usepackage{xspace}
\usepackage{amsmath}
\usepackage{multirow}
\usepackage{amsfonts}
\usepackage{enumitem}
\setlist[itemize]{leftmargin=*}
\setlist[enumerate]{leftmargin=*}
\usepackage{algorithm}
\usepackage{algorithmic}
\usepackage{amsthm}
\usepackage{url}
\usepackage{bm}
\usepackage{subfigure}
\usepackage{colortbl}
\definecolor{Gray}{gray}{0.85}

\theoremstyle{definition}
\newcommand{\mname}{\texttt{PopNet}\xspace}

\newcommand{\method}{\mname}

\newcommand{\tabincell}[2]{\begin{tabular}{@{}#1@{}}#2\end{tabular}}

\theoremstyle{definition}
\newtheorem{definition}{Definition}
\newtheorem{problem}{Problem}

\AtBeginDocument{%
  \providecommand\BibTeX{{%
    \normalfont B\kern-0.5em{\scshape i\kern-0.25em b}\kern-0.8em\TeX}}}

\copyrightyear{2022} 
\acmYear{2022} 
\setcopyright{acmcopyright}\acmConference[WWW '22]{Proceedings of the ACM Web
Conference 2022}{April 25--29, 2022}{Virtual Event, Lyon, France}
\acmBooktitle{Proceedings of the ACM Web Conference 2022 (WWW '22), April
25--29, 2022, Virtual Event, Lyon, France}
\acmPrice{15.00}
\acmDOI{10.1145/3485447.3512127}
\acmISBN{978-1-4503-9096-5/22/04}

\begin{document}
\title{\mname: Real-Time Population-Level Disease Prediction with Data Latency}





 \author{Junyi Gao}
\email{junyii.gao@gmail.com}
\affiliation{%
    \institution{IQVIA}
    \country{China}
}

\author{Cao Xiao}
\email{danica.xiao@amplitude.com}
\affiliation{%
    \institution{Amplitude}
    \country{USA}
}

\author{Lucas M. Glass}
\email{lucas.glass@iqvia.com}
\affiliation{%
    \institution{IQVIA}
    \country{USA}
}

\author{Jimeng Sun}
\email{jimeng@illinois.edu}
\affiliation{%
\institution{Department of Computer Science}
    \institution{University of Illinois Urbana-Champaign}
    \country{USA}
}


\begin{abstract}
Population-level disease prediction estimates the number of potential patients of particular diseases in some location at a future time based on (frequently updated) historical disease statistics. Existing approaches often assume the existing disease statistics are reliable and will not change. However, in practice,  data collection is often time-consuming and has time delays, with both historical and current disease statistics being updated continuously. In this work, we propose a real-time population-level disease prediction model which captures data latency (\mname) and incorporates the updated data for improved predictions. To achieve this goal, \mname models real-time data and updated data using two separate systems, each capturing spatial and temporal effects using hybrid graph attention networks and recurrent neural networks. \mname then fuses the two systems using both spatial and temporal latency-aware attentions in an end-to-end manner. We evaluate \mname on real-world disease datasets and show that \mname consistently outperforms all baseline disease prediction and general spatial-temporal prediction models, achieving up to 47\% lower root mean squared error and 24\% lower mean absolute error compared with the best baselines.
\end{abstract}



\maketitle

\section{Introduction}

Population-level disease prediction is of great significance to society since early forecasting of new disease counts at each location can help government or healthcare providers better optimize medical resources~\cite{Qian2020-ug} or inform where to build clinical trial sites~\cite{siettos2013mathematical,gao2020stan}. Compared with individual-level disease prediction, which predicts disease risk for each patient based on their health records~\cite{gao2019camp,ma2017dipole, choi2016doctor, choi2017using, gao2020stagenet, gao2020dr}, population-level disease prediction is usually based on frequently updated online historical disease statistics data collected from certain locations or population groups~\cite{gao2020stan,deng2020cola}.

Many machine learning or deep learning models have been developed to leverage patient data for individual disease prediction~\cite{gao2019camp, ma2017dipole, choi2016doctor, choi2017using, gao2020stagenet,choi2017gram,choi2018mime}. However, they cannot be applied to population-level disease prediction due to the need for accessing individual patient data. Meanwhile, existing population-level prediction models are mostly developed for infectious diseases. For example, epidemiology models such as the Susceptible-Infectious-Recovered (SIR) model were proposed for population-level infectious disease prediction~\cite{yang2020modified,kermack1927contribution,pei2020initial}. Recently, several works further proposed to augment such epidemiology models with deep neural networks for capturing spatial and temporal patterns~\cite{gao2020stan,deng2020cola}.

Existing population-level prediction models often assume that their model inputs (e.g., historical disease statistics) are reliable and accurate, which is often not true. In practice, data collection is time-consuming and has time delays, thus disease statistics require continuous updates to become more accurate~\cite{verrall2016understanding,clegg2002impact,song2012improved}. Such a data latency issue needs to be considered in population-level predictions. However, tackling this issue is not straight-forward. There are two main challenges:
\begin{itemize}[leftmargin=*]
    \item \textbf{Incorporating the updated data into the real-time model}. From the temporal perspective, a real-time model needs to be updated whenever an update is made to the historical data. From spatial perspective, different locations may be updated at different frequencies. We need to handle these  idiosyncratic updates in our model.
    \item \textbf{Extracting data updating patterns}. Data latency could be induced by various reasons, for example, geographic and demographic proximity between different locations, which causes the complexity of data updating patterns and brings difficulty for the model to utilize and make predictions. The noise and spatial-temporal correlation of different data streams also add to the difficulties of extracting data updating patterns.
   
\end{itemize}

To address these challenges, we propose a population-level disease prediction model (\mname) which captures data latency and incorporates the updated data for improved predictions. \mname is enabled by the following technical contributions.

\begin{itemize}[leftmargin=*]
 \item \textbf{Dual data modeling systems to incorporate updated data into the real-time model}. \mname models real-time data and updated data using two separate systems, each capturing spatial and temporal effects using hybrid graph attention networks (GAT) and recurrent neural networks (RNN). \mname then adaptively fuses the two systems using both spatial and temporal latency-aware cross-graph attentions in an end-to-end manner. To the best of our knowledge, we are the first work to incorporate updated data in spatio-temporal models.
 \item \textbf{Extract data updating patterns to enrich the spatial and temporal latency-aware attention}. We identify three major data updating patterns. (1) \textit{Spatial Correlation}. Geographically close locations may have similar data updating patterns and locations with similar populations may also have similar characteristics~\cite{gao2020stan,murray2015global}; (2) \textit{Seasonality}. The data updating patterns may be temporally periodic, and (3) \textit{Disease Correlation}. Disease comorbidities may lead to similar updating patterns. We enrich the spatial and temporal latency-aware attentions with these patterns, allowing the model to incorporate these patterns adaptively. 
\item \textbf{Efficient model update}.  \method can be trained efficiently on the newly added data via better initialization for hidden states of RNN. As a result, \mname can utilize previous historical patterns without reprocessing old data, which improves efficiency when the training sequences are long and brings convenience for deployment in real-world healthcare systems.
\end{itemize}

We evaluate \mname on real-world online medical claims datasets with real-time and update records and a  simulated synthetic dataset. Compared to the best baseline model, \mname achieves up to 47\% lower root mean squared error (RMSE) and 24\% lower mean absolute error (MAE) on two real-world disease prediction tasks.

\section{Related Works}

Over the years, spatial-temporal prediction models have been developed for application tasks such as traffic prediction~\cite{yao2018modeling,zheng2020gman,guo2019attention,huang2020lsgcn}, disease prediction~\cite{deng2020cola,gao2020stan,kapoor2020examining}, regional demand prediction~\cite{yao2018modeling} and general time-series prediction~\cite{cao2021spectral}. The recent success of deep learning models, especially GNNs and RNNs, brings promises to better model complex spatial and temporal features. Many research combines graph structures with disease statistics to model regional and temporal disease propagation and achieves more accurate predictions. For example, Deng et al.~\cite{deng2020cola} proposed a location attention mechanism and a graph message passing framework to predict influenza-like illness for different locations. Gao et al.~\cite{gao2020stan} incorporated clinical claims data in graph attention network to predict COVID-19 pandemics and use disease transmission dynamics to regularize RNN predictions. These models achieve good performance on their well-collected datasets. Compared with general spatio-temporal prediction works, our work more focuses on incorporating updated data into the spatio-temporal model. Since in practice, the input data is not always reliable due to latency or errors and may get updated in the future. We believe this scenario is common in web data and real-world settings.

Consider broader spatial-temporal prediction models in other fields such as traffic prediction, most works also utilize graph neural networks to extract spatial features and use RNNs or attention mechanisms to extract temporal features~\cite{xu2020inductive,trivedi2019dyrep,wang2021inductive,kumar2019predicting,pareja2020evolvegcn}. Those works also do not have the consideration or model design for data latency. For example, the traffic prediction model GMAN~\cite{zheng2020gman} leverages the node2vec approach to preserve graph structure in node embeddings and then samples the neighboring nodes to obtain the embedding. Guo et al.~\cite{guo2019attention} proposed ASTGCN to extract multi-scale temporal features by training three network branches to receive hour-level, day-level, and week-level data. In our work, we enrich the model with spatial and temporal background information, making the model adaptively extract spatial relationships of both close nodes and distant but similar nodes, also from multiple time scales. 


\section{Problem Formulation}


\begin{definition}[\textbf{Disease statistics data}]
The disease statistics data are collected from medical claims or online reports of local health departments from different locations. They can be represented as a 3D tensor $\mathcal{X}\in \mathbb{R}^{N\times T\times F}$, where $N$ denotes the number of locations, $T$ is the number of total timesteps, $F$ is the number of features (i.e., diseases). Matrix $\mathbf{X}^{t}\in\mathbb{R}^{N\times F}$ and $\mathbf{X}_{i}\in\mathbb{R}^{T\times F}$ denote slices from the $\mathcal{X}$ tensor from time dimension and location dimension. Vector $\mathbf{x}_i^{t}\in\mathbb{R}^{F}$ denotes a slice from the $\mathbf{X}^{t}$ matrix at $i$-th location.
\end{definition}

\begin{definition}[\textbf{Updated disease data}]
The real-time statistics maybe unreliable due to time delays during data collection process, therefore every tensor element in $\mathcal{X}$ may be updated at a future timestep. For example, for a specific location at timestep $t$, after we obtain the initial disease statistics for this timestep, we may constantly receive updates for the statistics of timestep $t$ in future timesteps $t+1$, $t+2$, …. All the updated values consist of the updated disease data $\mathcal{U}\in \mathbb{R}^{N\times T\times F}$, which is a 3D tensor. Similar to the original disease data, we also use $\mathbf{U}^{t}\in\mathbb{R}^{N\times F}$, $\mathbf{U}_{i}\in\mathbb{R}^{T\times F}$ and $\mathbf{u}_i^{t}\in\mathbb{R}^{F}$ to denote different slices from the updated data tensor. Here $\mathbf{u}_i^{t_1}$ refers to data updated for location $i$ at a future time $t_1$. Suppose it will replace the original disease data $\mathbf{x}_i^{t_0}$, note that $t_1>t_0$, we define this update latency as $\Delta t_i=t_1-t_0$. All update latency is aggregated to a 2D matrix $\Delta\mathbf{t}\in \mathbb{R}^{N\times T}$. Value $0$ in $\mathcal{U}$  means no updates for those tensor elements.
\end{definition}

\begin{definition}[\textbf{Location graph}]
A location graph can be modeled as an undirected graph $\mathcal{G}=(V, E,\mathbf{A})$, where $\mathbf{V}$ is the set of $|V|=N$ location nodes, $E$ is the set of edges, $\mathbf{A}$ denotes the adjacency matrix of the graph. The edges are computed based on the geographical and demographic proximity between locations, which will be detailedly introduced in following sections.
\end{definition}

\begin{problem}[\textbf{Spatial-temporal disease prediction}]
Given historical original disease statistics $\mathcal{X}$ and updated disease data $\mathcal{U}$, 
the population-level spatial-temporal disease prediction task is a regression task, which is to predict the future ground-truth number of cases for a certain disease $\mathbf{Y}\in \mathbb{R}^{N}$ for all $N$ locations at $T+1$ timestep. We also support multiple-step prediction for next $k$ steps from $T+1$ to $T+k$ timesteps.
\end{problem}

\begin{figure*}[h!]
    \centering
    \includegraphics[width=0.9\textwidth]{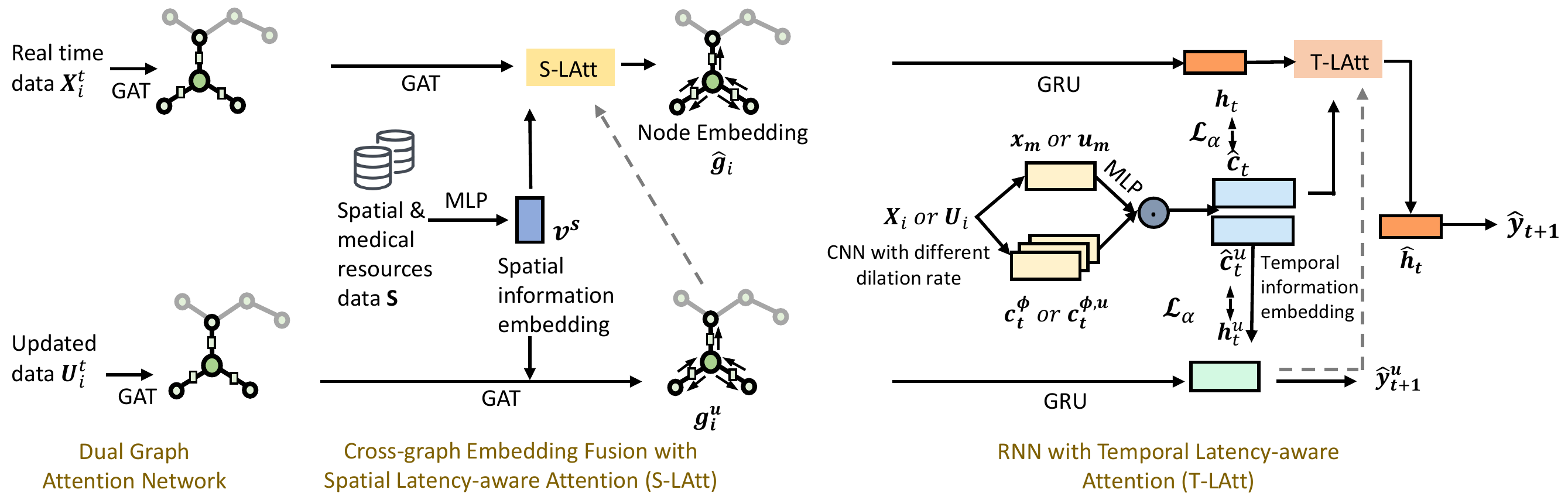}
    \caption{
   Our \mname model consists of two graph attention networks to receive real-time data ($\mathcal{X}$) and updated data ($\mathcal{U}$) respectively. It uses spatial latency-aware attention (S-LAtt) to fuse two graphs and generate node embedding for each location. The spatial latency-aware attention is enriched by spatial information embedding (SIE) $\mathbf{v}^s$ learned  using location-wise geographical and medical resource features. The node embeddings in two graph networks are fed into two GRUs respectively to extract temporal relations. \mname also utilizes temporal latency-aware attention (T-LAtt) to fuse temporal embeddings. Similarly, T-LAtt is enriched by temporal information embeddings (TIE) $\hat{\mathbf{c}}_t$ and $\mathbf{c}_t^u$, which can adaptively embed the most informative multi-scale disease patterns to improve predictions. \mname also aligns the hidden states of GRU $\hat{\mathbf{h}}_t, \mathbf{h}_t^u$ with the learned TIE $\hat{\mathbf{c}}_t$ and $\mathbf{c}_t^u$ respectively to achieve efficient update. Finally, \mname will output predictions $\hat{y}_{t+1}$ using fused temporal embedding.}
    \label{fig:model}
\end{figure*}

\section{The \mname Model}

As shown in Fig.~\ref{fig:model},   \mname models real-time data $\mathcal{X}$ and updated data $\mathcal{U}$ using two separate systems, and then adaptively fuses the two systems using both spatial and temporal latency-aware cross-graph attention. Below we introduce \mname in more details.

\subsection{Dual Graph Attention Network}
We model the real-time data $\mathcal{X}$ and updated data $\mathcal{U}$ using separate systems for better modal capacities of pattern extractions in both data sources. We employ graph attention networks (GAT)~\cite{velivckovic2017graph} to leverage spatial relations between locations. This way, the prediction for a target location can be improved by utilizing spatial disease patterns discovered in nearby or similar locations. We use two graph attention networks to process $\mathcal{X}$ and $\mathcal{U}$ respectively.

Here $\mathcal{X}$ and $\mathcal{U}$ share the same undirected graph design $\mathcal{G}(V, E, \mathbf{A})$. In graph $\mathcal{G}$, the nodes indicate locations; while the edge connecting node $i$ and $j$, denoted as $w_{ij}$, is the similarity between node $i$ and $j$ such that $w_{ij} = p_i^\alpha p_j^\beta exp(-\frac{d_{ij}}{\gamma})$. Here $d_{ij}$ is distance between node $i$ and $j$, $p_i$ is the population size of node $i$, and $\alpha, \beta, \gamma$ are hyper-parameters. We use a threshold value $\omega$ to calculate the graph adjacency matrix as in Eq.~\eqref{eq:graphadj},
\begin{equation}
\left\{
             \begin{array}{lr}
             A_{ij} = 1,\,if\,w_{ij} \geq \omega \\
             A_{ij} = 0,\,if\,w_{ij} < \omega \\
             \end{array}
\right.
\label{eq:graphadj}
\end{equation}
For simplicity,  we focus our discussion in this section on a specific timestep and thus will omit the superscript $t$. Accordingly the attention score between node $i$ and $j$ will be computed as in Eq.~\eqref{eq:newattention},
\begin{equation}
\begin{aligned}
    &\mathbf{z}_i = \mathbf{W}_z\mathbf{x}_i,\;\mathbf{z}_i^u = \mathbf{W}_z^u\mathbf{u}_i\\
    &e_{ij} = \sigma(\mathbf{W}_a(\mathbf{z}_i|\mathbf{z}_j)),\;e_{ij}^u = \sigma(\mathbf{W}_a^u(\mathbf{z}_i^u|\mathbf{z}_j^u))
\end{aligned}
\label{eq:newattention}
\end{equation}
where $\mathbf{W}_z\in\mathbb{R}^{|\mathbf{z}_i|\times F}$, $\mathbf{W}_a\in\mathbb{R}^{|\mathbf{z}_i|+|\mathbf{z}_j|}$, $\mathbf{W}_z^u\in\mathbb{R}^{|\mathbf{z}_i^u|\times F}$, $\mathbf{W}_a^u\in\mathbb{R}^{|\mathbf{z}_i^u|+|\mathbf{z}_j^u|}$ are attention weight matrices for the GAT networks, $\sigma$ denotes the LeakyReLU activation function, and $(\cdot|\cdot)$ denotes the concatenate operation. Then we use softmax function to normalize the obtained attention score as in Eq.~\eqref{eq:attentionscore},
\begin{equation}
    a_{ij} = \frac{\exp{(e_{ij})}}{\sum_{k\in \mathcal{N}(i)}{\exp{(e_{ik})}}},\quad a_{ij}^u = \frac{\exp{(e_{ij}^u)}}{\sum_{k\in \mathcal{N}(i)}{\exp{(e_{ik}^u)}}}
    \label{eq:attentionscore}
\end{equation}
where $\mathcal{N}(i)$ denotes the set of one-hop neighbors of node $i$.

Likewise, we use the multi-head attention mechanism~\cite{velivckovic2017graph} to enrich the model capacity by calculating $K$ independent attention scores, where $K$ is the number of attention heads. We obtain the aggregated node embedding  as given by Eq.~\eqref{eq.initial_node},
\begin{equation}
    \mathbf{g}_{i} = \sigma(\frac{1}{K}\sum_{k=1}^{K}{\sum_{j\in \mathcal{N}(i)}{a_{ij}^k\mathbf{W}^{k}_{g}\mathbf{x}_j}}), \mathbf{g}_{i}^u = \sigma(\frac{1}{K}\sum_{k=1}^{K}{\sum_{j\in \mathcal{N}(i)}{a_{ij}^{u,k}\mathbf{W}^{u,k}_{g}\mathbf{u}_j}})
\label{eq.initial_node}
\end{equation}
where $\mathbf{W}^{k}_{g}\in\mathbb{R}^{|\mathbf{g}_i|\times F}$ and $\mathbf{W}^{u,k}_{g}\in\mathbb{R}^{|\mathbf{g}_i^u|\times F}$ are the weight matrices for the $k$-th attention head in two GATs, respectively. Therefore, for each node $i$, we will obtain two node embeddings $\mathbf{g}_{i}$ and $\mathbf{g}_i^u$ for real-time data $\mathcal{X}$ and updated data $\mathcal{U}$ respectively.

\subsection{Cross-Graph Embedding Fusion with Spatial Latency-aware Attention}
After obtaining all the node embeddings for updated data and real-time data, we would like to utilize the updated historical data to make better predictions. However, this is not a straightforward task. The latency in data updating can vary between two embeddings of the same node. Hence, directly concatenating or summing two embeddings may confuse the prediction network and lead to inferior prediction results. Besides, there is a latency in the updated data because those locations can be updated at a different frequency. 
To incorporate these complex latency patterns, we design the spatial latency-aware attention (S-LAtt) to fuse spatial embeddings.

The idea of S-LAtt is to use the node embedding as the query to aggregate spatial patterns from  nearby or similar nodes (i.e., locations), assuming they have similar data updating patterns. To better quantify such similarity, we learn a spatial information embedding (SIE) $\mathbf{v}^{s}$ for each node, where the spatial information includes populations, the numbers of hospitals and ICU beds, longitude, and latitude. For node $i$, SIE is obtained via Eq.~\eqref{eq:siemlp},
\begin{equation}
    \mathbf{v}_{i}^{s} = MLP(\mathbf{S}_i)
    \label{eq:siemlp}
\end{equation}
where $\mathbf{S}_i$ denotes the spatial information of node $i$. Since these spatial information are in general  static, same nodes in both GATs share the same $\mathbf{S}_i$. Based on the  the node embedding and the SIE, we compute the cross-graph attention score as in Eq.~\eqref{eq:crossgraphattention},
\begin{equation}
    e_{ij} = \sigma(\mathbf{W}_a(\mathbf{W}_g(\mathbf{g}_i|\mathbf{v}_{i}^{s}) + \mathbf{W}_u(\mathbf{g}^u_{j} | \mathbf{v}_{j}^{s})))
    \label{eq:crossgraphattention}
\end{equation}
where $j\in \mathcal{N}(i)$, $\mathcal{N}(i)$ is the neighboring nodes set of node $i$ in the location graph $\mathcal{G}$.

In addition to spatial similarity, we also notice that the longer the latency is, the smaller the marginal influence the new data will have on our final prediction, thus we use time latency to regularize this attention score. To be specific, we utilize the temporal latency $\Delta t_{ij}$ between node $i$ and $j$, and design a heuristic function to use this temporal latency $\Delta t$ as in Eq.~\eqref{eq:deltaregularize},
\begin{equation}
    f(\Delta t_{ij}) = \frac{1}{\log{(1+\exp{(\Delta t_{ij})})}}.
    \label{eq:deltaregularize}
\end{equation}
Then the attention weight $a_{ij}$ is regularized  as in Eq.~\eqref{eq:regularize},
\begin{equation}
   \hat{e}_{ij} = e_{ij}  f(\Delta t_{ij}), \quad
   a_{ij} = \frac{\exp{(\hat{e}_{ij})}}{\sum_{k\in \mathcal{N}_u(i)}{\exp{(\hat{e}_{ik})}}},
\label{eq:regularize}
\end{equation}
and we  can get the aggregated updated embedding as in Eq~\eqref{eq.agg_node}.
\begin{equation}
    \mathbf{\hat{g}}^u_{i} = \sigma(\sum_{j\in \mathcal{N}_u(i)}{a_{ij}\mathbf{g}^u_{j}})
    \label{eq.agg_node}
\end{equation}
Finally, we concatenate the node embedding $\mathbf{g}_i$, the aggregated updated embedding $\mathbf{\hat{g}}^u_{i}$ and the original input data $\mathbf{x}_i$ as in Eq.~\eqref{eq.final_node}.
\begin{equation}
   \mathbf{\hat{g}}_i = (\mathbf{g}_i | \mathbf{\hat{g}}^u_{i} | \mathbf{x}_i)
   \label{eq.final_node}
\end{equation}

\subsection{Recurrent Neural Network with Temporal Latency-aware Attention}
In addition to spatial patterns, we also employ the gated recurrent unit networks~\cite{chung2014empirical} to extract the temporal patterns based on multivariate time series from each node. To simplify, we focus our discussion on one location and 
omit the subscript node index $i$. The real-time and updated embeddings are fed into GRUs as in Eq.~\eqref{eq.init_ht},
\begin{equation}
\begin{aligned}
   & \mathbf{h}_{t} = GRU(\mathbf{\hat{g}}_1, \mathbf{\hat{g}}_2, ..., \mathbf{\hat{g}}_t)\\
    &\mathbf{h}_{t}^u = GRU_u(\mathbf{g}^u_1, \mathbf{g}^u_2, ..., \mathbf{g}^u_t)
    \end{aligned}
    \label{eq.init_ht}
\end{equation}
We use the hidden states of the GRU $\mathbf{h}_{t}$ and $\mathbf{h}_{t}^u$ as the temporal embeddings. Similarly, we design a temporal latency-aware attention mechanism (T-LAtt) to fuse two embeddings and deal with the latency between $\mathbf{h}_{t}$ and $\mathbf{h}_{t}^u$. As previously discussed, utilizing temporal-related data updating patterns may benefit the predictions. This involves extracting complex temporal patterns such as increasing or declining from different time scales.  Besides, we also consider the updating patterns of comorbidities of target diseases. To extract and leverage these patterns, we enrich the T-LAtt with temporal information embeddings (TIE).

First, for  TIE to extract temporal patterns from multiple timescales, we use dilated convolutional networks~\cite{oord2016wavenet} with different dilation rates to extract  temporal patterns from different time scales. Concretely, at each location, the input disease data sequence $\mathbf{X} = [\mathbf{x}^1, \mathbf{x}^2, ..., \mathbf{x}^t$] is fed into the CNN as in Eq.~\eqref{eq:multiscale},
\begin{equation}
    \mathbf{c}_t^\phi = \mathbf{m}(L, \phi) * \mathbf{X},
    \label{eq:multiscale}
\end{equation}
where $*$ denotes the convolution operation, $\mathbf{m}(L, \phi)$ is the 1D convolution filter with size $L$ and dilation rate $\phi$. The larger the $\phi$ is, the larger the filter's receptive field is, making the convolution filter extract temporal patterns from a broader time scale. In our experiments, we use a combination of different $\phi$ to extract patterns in different scales, from small to large. The feature maps are concatenated to get the final feature map vector $\mathbf{c}_t\in \mathbb{R}^{C}$, $C$ denotes the number of convolution filters. Each value in $\mathbf{c}_t$ represents an extracted temporal feature. 

Following previous CNN-based models~\cite{gao2020stagenet,hu2018squeeze,ma2020adacare}, we also try to select the most informative patterns in $\mathbf{c}_t$ based on attention weights. Here, we first use mean pooling over time dimension for $\mathbf{X}$ as $\mathbf{x}_m = MeanPool(\mathbf{X})$, $\mathbf{x}_m \in \mathbb{R}^F$. $\mathbf{x}_m$ can be regarded as a summary for $F$ diseases. This vector is used to calculate the attention score for the temporal patterns as in Eq.~\eqref{eq:vector},
\begin{equation}
    \mathbf{a}_t^c = \sigma(MLP(\mathbf{x}_m))
    \label{eq:vector}
\end{equation}
where $\sigma$ denotes the sigmoid activation. We use the multi-layer perceptron to do the mapping $\mathbb{R}^{F} \rightarrow \mathbb{R}^{C}$, and the sigmoid activation to generate importance score between 0 and 1. The obtained score vector $\mathbf{a}_c$ is used to re-calibrate the feature map vector as in Eq.~\eqref{eq.tie},
\begin{equation}
    \mathbf{\hat{c}}_t = \mathbf{c}_t \odot \mathbf{a}_t^c
    \label{eq.tie}
\end{equation}
The obtained $\mathbf{\hat{c}}_t$ is the final temporal information embedding (TIE). We can also get the TIE for updated series $\mathbf{\hat{c}}_t^{u}$ in this way.

Similar to the spatial latency-aware attention, we use the TIE to enrich the attention and use time latency between current temporal embedding $\mathbf{h}_t$ and historical updated temporal embeddings $[\mathbf{h}_1^u, \mathbf{h}_2^u, ..., \mathbf{h}_t^u]$ to regularize the attention score as in Eq.~\eqref{eq:tiescore},
\begin{equation}
\begin{aligned}
   & e_{ti} = f(\Delta t_{ti}) * \sigma(\mathbf{W}_a(\mathbf{W}_{h1}(\mathbf{h}_t|\mathbf{\hat{c}}_t) + \mathbf{W}_{h2}(\mathbf{h}^u_{i} | \mathbf{\hat{c}}^{u}_t)))\\
   & a_{ti} = \frac{\exp{(e}_{ti})}{\sum_{j=1}^{t}{\exp{(e_{tj})}}}
   \end{aligned}
   \label{eq:tiescore}
\end{equation}
And the aggregated updated temporal embedding is given by Eq.~\eqref{eq.agg_ht},
\begin{equation}
    \mathbf{\hat{h}}_t^u = \sum_{i=1}^{t}{a_{ti}\mathbf{h}^u_{i}}.
    \label{eq.agg_ht}
\end{equation}
Finally, we concatenate the aggregated updated temporal embedding $\mathbf{\hat{h}}_t^u$, the original temporal embedding and the TIE to calculate the final temporal embedding as in Eq.~\eqref{eq.final_ht},
\begin{equation}
    \mathbf{\hat{h}}_t = (\mathbf{h}_t | \mathbf{\hat{h}}^u_{t} | \mathbf{\hat{c}}_t).
    \label{eq.final_ht}
\end{equation}

\subsection{Efficient Iterative Training and Prediction}
\noindent{\bf Model update challenge:} In clinical practice, an online population-level disease prediction model needs routine updates when new data become available. 
For model updating with new data, it usually requires model retraining which is time-consuming as the data sequence becomes longer, or directly fine-tuning on the new data which discards historical patterns. Some also  initialize a  model using new data and the last hidden state of RNN trained using old data. However, this solution assumes there is no large time gap between the two datasets. Otherwise, capturing the continuous behavior of the system becomes more difficult for the model and leads to even worse performance~\cite{mohajerin2017state}. \\

\noindent{\bf Our Solution}.  \mname introduces an alignment module to address this issue via providing a better initialization for the hidden states of the RNN on the new data without assuming the data continuity. This is achieved by learning a mapping function between the TIE $\hat{\mathbf{c}}_t$, $\mathbf{c}_t^u$ and the hidden states of RNN $\hat{\bm{h}}_t$ and $\mathbf{h}_t^u$ at each timestep respectively. When applying to new data, directly using the last hidden states is not optimal due to the new disease patterns may be different. But convolutional features can provide a better initialization for the RNN since they are not strictly sequential-dependent. 
Concretely, we first use a mapping function $m_\theta$ parameterized by $\theta$ to map the learned TIE to another latent space as in Eq.~\eqref{eq.map},
\begin{equation}
    \hat{\bm{c}}_t^m = m_\theta(\hat{\bm{c}}_t)
    \label{eq.map}
\end{equation}
Then we calculate the probability distribution of the mapped TIE embedding and the current hidden state of the RNN $\bm{h}_t$ using softmax function as in Eq.~\eqref{eq.dist},
\begin{equation}
    p(\hat{\bm{c}}_t^m) = softmax(\hat{\bm{c}}_t^m); \quad q(\bm{h}_t) = softmax(\bm{h}_t)
    \label{eq.dist}
\end{equation}

Then we define the alignment loss function between $p(\hat{\bm{c}}_t^m)$ and $q(\bm{h}_t)$ using Kullback-Leibler divergence as in Eq.~\eqref{eq.la},
\begin{equation}
     \mathcal{L}_a = \sum_{i=1}^n{p(\hat{\bm{c}}_t^m)}\log(\frac{p(\hat{\bm{c}}_t^m)}{q(\bm{h}_t)})
     \label{eq.la}
\end{equation}

Note that here we use the Kullback-Leibler divergence since its asymmetric characteristic naturally fits our design: we expect the loss term can help the model learn a close estimation to $\bm{h}_t$ using $\hat{\bm{c}}_t$. Besides, since the dimensionality of two embeddings is large, using KL divergence instead of L1 or L2 distance can also help avoid the learned $m_\theta$ simply mapping the embeddings to random normal distributions. 
When applied to new data, the model will first calculate the TIE using the entire sequence and then use $m_\theta$ to provide the initialization for the hidden states of the RNN. The detailed algorithm is shown in Alg.\ref{alg:Framwork}.

Finally, we use a two-layer perceptron to generate predictions via $\hat{y}_{t+1} = MLP(\mathbf{\hat{h}}_t)$. We also let the $GRU_u$ to make predictions as $\hat{y}^{u}_{t+1} = MLP(\mathbf{h}^u_t)$. Note that $\hat{y}^{u}_{t+1}$ is the prediction for the day after the update point, so it may be earlier than current timestep $t$. However, we use this as an auxiliary task to better optimize the GRU and GAT. At testing time, only $\hat{y}_{t+1}$ is the model output. We use mean squared error as the loss function as in Eq.~\eqref{eq.lp},
\begin{equation}
        \mathcal{L}_r = \frac{1}{n}\sum_{i=1}^{n}{(\hat{y}_{i+1} - y_{i+1})^2}; \quad
        \mathcal{L}_u = \frac{1}{n}\sum_{i=1}^{n}{(\hat{y}_{i+1}^u - y_{i+1}^u)^2} 
        \label{eq.lp}
\end{equation}
We finally optimize the entire model using Eq.~\eqref{eq.l}.
\begin{equation}
    \mathcal{L} = \mathcal{L}_r + \mathcal{L}_u + \mathcal{L}_a
    \label{eq.l}
\end{equation}

\begin{algorithm}[h!] 
\caption{The \mname model} 
\label{alg:Framwork} 
\begin{algorithmic}
\REQUIRE ~~\\ 
Real-time disease statistics $\mathcal{X}$, updated disease statistics $\mathcal{U}$, update intervals $\bm{\Delta t} = [\bm{\Delta t}_{1}, \bm{\Delta t}_{2}, ..., \bm{\Delta t}_{T}]$, prediction targets $\bm{Y} = [\bm{Y}_{1}, \bm{Y}_{2}, ..., \bm{Y}_{T}]$ and location graph $\mathcal{G}$. \\
\ENSURE ~~\\ 
\FOR{$i=1$ to $T$}
\STATE Input $\mathcal{X}$ and $\mathcal{U}$ and get node embeddings using Eq.~\ref{eq.initial_node}; \\ 
\STATE Aggregate $\hat{\mathbf{g}}^u$ to $\hat{\mathbf{g}}$ using Eq.~\ref{eq.agg_node} and \ref{eq.final_node};\\
\STATE Input spatial embeddings to GRU networks using Eq.~\ref{eq.init_ht}; \\
\STATE Calculate temporal information embeddings using Eq.~\ref{eq.tie}; \\
\STATE Aggregate $\mathbf{h}^u_i$ to $\mathbf{h}_i$ using Eq.~\ref{eq.agg_ht} and \ref{eq.final_ht}; \\
\STATE Make predictions for $i+1$ timestep;\\
\ENDFOR \\
\STATE Generate distributions using Eq.~\ref{eq.map} and \ref{eq.dist}; \\
\STATE Calculate KL divergence using Eq.~\ref{eq.la}; \\
\STATE Optimize model parameters by minimizing loss in Eq.~\ref{eq.l}. \\
\RETURN ~~\\
\STATE Calculate TIE for the input sequence using Eq.~\ref{eq.tie}; \\
\STATE Use learned $m_\theta$ to generate the initial hidden state of GRU; \\
\STATE Repeat the normal training steps. \\
\end{algorithmic}
\end{algorithm}

\section{Experiment}

We evaluate \mname by comparing against several spatial-temporal prediction and disease prediction baselines using real-world datasets.

\subsection{Experimental Setup}

\noindent\textbf{Data}
We extract disease statistics from patients' claims data in a real-world patient database from IQVIA.
The patients' claims data are collected from 2952 counties in the US starting from 2018. We aggregate the ICD-10 codes in claims data into 21 categories, which include 17 diseases and 4 other codes (see detailed category descriptions in Appendix). We use week-level statistics. Since patients' claims data cannot be completed collected at one time, the disease statistics in a certain week will be updated over several weeks. We use the statistics collected in the first week as the real-time data, and we use the data collected from future weeks as the updated data. We conduct experiments to predict two diseases:

\begin{enumerate}
    \item \textbf{Respiratory Disease Dataset}: Respiratory diseases include ICD10 codes J00-J99, which are common and most of them are contagious. The number of cases is larger than most other diseases. Therefore, the claims data collection procedure is also longer, so that the disease statistics for one week will be fully collected in the following up to 13 weeks. We filter out locations that have very few cases (less than 100). Finally, we get 1,693 counties for respiratory diseases prediction.
    \item \textbf{Tumors Dataset}: Tumors include ICD10 codes C00-D49. Compared to respiratory diseases, the tumors have fewer cases, and the data update period is also shorter. Most statistics of one week can be fully collected in the following 7 weeks. We also filter out locations with very few cases (less than 10), and we get 1,829 counties for tumor prediction.
\end{enumerate}
In addition, we conduct experiments on other 15 diseases and a synthetic dataset generated based on real-world disease update distributions. The detailed statistics and the results can  be found in the Appendix. The code and the synthetic dataset is publicly available in~\footnote{https://github.com/v1xerunt/PopNet}.\\

\noindent\textbf{Baselines}
We evaluated \mname against the following spatio-temporal prediction baselines: \textbf{SARIMAX}, \textbf{GRU}~\cite{chung2014empirical}, \textbf{ASTGCN}~\cite{guo2019attention}, \textbf{GMAN}~\cite{zheng2020gman}, \textbf{EvolveGCN}, \textbf{ColaGNN}~\cite{deng2020cola} and \textbf{STAN}~\cite{gao2020stan}. The detailed descriptions of baselines can be found in Appendix.
    
We also compare \mname with the reduced version as the ablation study.
\begin{enumerate}
    \item \textbf{\mname-LAtt} We reduce both S-LAtt and T-LAtt mechanisms from \mname. \mname-LAtt is essentially two branches that receive real-time and updated data independently, and the outputs of two networks are concatenated to make final predictions.
    \item \textbf{\mname-SLAtt} We only reduce the spatial latency-aware attention from \mname.
    \item \textbf{\mname-TLAtt} We only reduce the temporal latency-aware attention from \mname.
    \item \textbf{\mname-$\mathcal{L}_\alpha$} We reduce the alignment module and the loss term $\mathcal{L}_\alpha$ from \mname. Since this term is only related to the iterative training, it will only be evaluated  in Q3 section. \mname-$\mathcal{L}_\alpha$ simply uses normal initialization for  RNN hidden states.
\end{enumerate}

\noindent\textbf{Metrics}. Following the similar work~\cite{gao2020stan, velivckovic2017graph}, we use the following regression metrics to evaluate all the models:
The \textbf{root mean squared error (RMSE)}, \textbf{mean absolute error (MAE)} and the \textbf{mean absolute percentage error (MAPE)} measures the difference between predicted values and true values:
\begin{equation}
    RMSE = \sqrt{\frac{1}{n}\sum_{i=1}^{n}{(\hat{y}_i - y_i)^2}}
\end{equation}
\begin{equation}
    MAE = \frac{1}{n}\sum_{i=1}^{n}{|\hat{y}_i - y_i|}
\end{equation}
\begin{equation}
MAPE = \frac{1}{n}\sum_{i=1}^{n}{\frac{|\hat{y}_i - y_i|}{y_i}}
\end{equation}
All metrics are calculated after projecting the values into the real range.\\

\noindent\textbf{Evaluation Strategy}.
We split the data into training, validation, and testing sets. The training set is 60 weeks, starting from January 2018 to March 2019. The validation set is 20 weeks, starting from April 2019 to July 2019. The testing set is 20 weeks, starting from August 2019 to December 2019. All the models use the same training data and are also evaluated and tested using the same sets. To be fair, the training, validation and test data for all models are the same. For baseline models, we concatenate the updated data at each time step with the real-time data for model inputs. Compared to all baselines, \mname does not access any extra data. In order to reduce the variance caused by time shift, we save three model checkpoints with (1) lowest loss on the training set; (2) lowest MSE on the validation set; (3) last training epoch. We test each checkpoint on the test set and report the best performance.\\

\noindent\textbf{Implementation Details}. All methods are implemented in PyTorch~\cite{paszke2019pytorch} and trained on an Ubuntu 16.04 with 64GB memory and a Tesla V100 GPU. We use Adam optimizer \cite{kingma2014adam} with a learning rate of 0.001 and trained for 200 epochs. The hyper-parameter settings of each baseline model can be found in the appendix.

\subsection{Results}






\subsection*{Q1. Performance on Disease Prediction}

The prediction results on respiratory disease and tumors  are listed in Table~\ref{tumor_res}. We also conduct two-tailed student’s T-test of MAE between \mname and other baseline models to test the significance of performance improvement. The p-values are also in Table \ref{tumor_res}.
\mname  outperforms all baseline methods on all metrics. On respiration disease dataset,  \mname achieves 47\% lower RMSE, 23.2\% lower MAE, and 29.4\% lower MAPE, and on tumors dataset, \mname achieves 29\% lower RMSE, 24\% lower MAE, and 13.2\% lower MAPE, both compared with the best baseline ColaGNN.

\begin{table}[h!]
    \caption{Disease Prediction Performance}
\begin{tabular}{lcccc}
\toprule
\multicolumn{5}{c}{\bf Respiratory Diseases Prediction}\\
                        \textbf{Model}   & \textbf{RMSE ($\times10^5$)} & \textbf{MAE} & \textbf{MAPE} & \textbf{P-value} \\ \hline
 SARIMAX             & 15.54 & 542.5 & 51.3 & 0.0 \\
                          GRU             & 10.89 & 340.2&43.2 & 5e-20 \\
                          GMAN             & 9.10 & 329.8&37.4 & 4e-15 \\
                        ASTGCN             & 10.33 & 303.6&39.9 & 4e-13 \\
                        EvolveGCN & 9.85 & 312.4 & 36.9 & 8e-9 \\
                          STAN             & 9.54 & 305.7&36.8 & 4e-9 \\
                          ColaGNN        & 8.06 &291.3&33.7 & 5e-5 \\ 
\rowcolor{Gray} \mname-LAtt&9.82          &    311.6  &    39.2 & 4e-10 \\
\rowcolor{Gray} \mname-TLAtt       &      6.32 &    271.3           &   29.9  &  5e-4   \\
                     \rowcolor{Gray}      \mname-SLAtt &      8.85 &    297.5   &    31.3  &  7e-8\\  
                     \rowcolor{Gray} \mname&\textbf{4.29}          &    \textbf{223.8}           &    \textbf{23.8} & - \\
\midrule
\multicolumn{5}{c}{\bf Tumors Prediction}\\
                           \textbf{Model}   & \textbf{RMSE ($\times10^5$)} & \textbf{MAE} & \textbf{MAPE} & \textbf{P-value} \\ \hline
 SARIMAX             & 21.41 & 426.0 &69.8&0.0 \\
                          GRU             & 16.20 & 313.3&60.7&0.0 \\
                          GMAN             & 13.12 & 315.7&53.4&7e-25 \\
                          ASTGCN             & 15.59 & 332.5&56.2&0.0 \\
                          EvolveGCN & 8.94 & 269.0 & 50.4 & 5e-11 \\
                         STAN             & 6.56 & 215.8&48.7&7e-9 \\
                        ColaGNN        &4.75&172.5&42.3&9e-5  \\ 
\rowcolor{Gray} \mname-LAtt      &      8.92          &    283.4          &    51.5 & 4e-10         \\
\rowcolor{Gray} \mname-TLAtt      &      3.25          &    142.9           &    38.1 &3e-4        \\
\rowcolor{Gray}     \mname-SLAtt        &  4.50          &    165.8           &    40.1    &  5e-5       \\ 
\rowcolor{Gray}  \mname&\textbf{2.90}          &    \textbf{131.6}           &    \textbf{36.7}&- \\
\bottomrule
\end{tabular}
\label{tumor_res}
\vskip -1em
\end{table}

\begin{figure}
    \centering
    \includegraphics[width=\columnwidth]{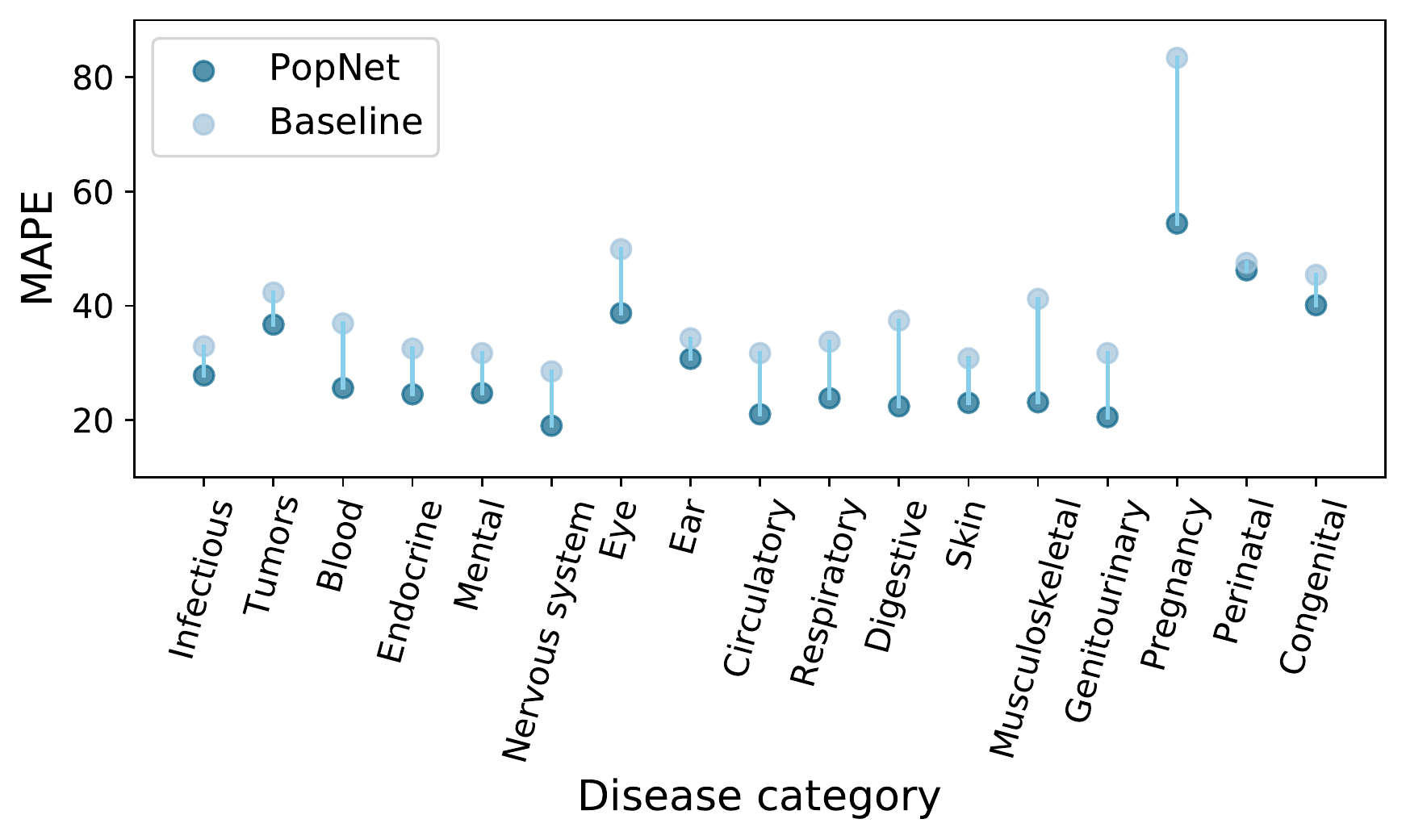}
    \caption{Comparison of the test MAPE between \mname and the best baseline on all disease code categories}
    \label{fig:mape}
    \vskip -1em
\end{figure}

In addition, we  evaluate the  performance on other disease categories (grouped by ICD code). We report the detailed test MAE in Appendix, and show the test MAPE of \mname against the best baseline in Fig.~\ref{fig:mape}.
The results show that \mname outperforms all baseline models for all disease categories, which indicates the potential broader utility of \mname. It is worth noting that for some disease codes, \mname achieves much better performance than baseline models, for example, musculoskeletal disease or pregnancy prediction. These data often receive more frequent updates due to the large number of cases or the particularity. Therefore, \mname achieves better performance since it can better extract and utilize update patterns.

\subsection*{Q2. Performance at Different Locations}
This section further explores the performance of \mname on different locations
We report the number of locations that each model has the best performance in Table.~\ref{gap_res}. Here 'Others' sums up the results of SARIMAX, GMAN, GRU, ASTGCN and EvolveGCN.

\begin{table}[h!]
    \caption{\# of locations where each model performs the best.}
\begin{tabular}{lccc}
\toprule
\multicolumn{4}{c}{\textbf{Respiratory Diseases}} \\
\textbf{Model}   & \# of Locations & \% of Locations & Mean $\Delta MAPE$    \\ \hline
STAN             & 42 & 2.5\% & 3.5\%  \\
ColaGNN        & 33 & 1.9\% & 2.7\%   \\
Others & 6 & 0.4\% & 4.1\%   \\ \rowcolor{Gray}
\mname& \textbf{1612}& \textbf{95.2\%} & \textbf{12.4\%}  \\
\midrule
\multicolumn{4}{c}{\textbf{Tumors}} \\
\textbf{Model}   & \# of Locations & \% of Locations & Mean $\Delta MAPE$    \\ \hline
STAN            & 40 & 2.2\% & 8.5\%  \\
ColaGNN       & 122 & 6.7\% & 4.6\%   \\
Others & 22 & 1.2\% & 5.5\%  \\ \rowcolor{Gray}
\mname& \textbf{1645}& \textbf{89.9\%} & \textbf{10.7\%}  \\
\midrule
\end{tabular}
\label{gap_res}
\end{table}
It is easy to see \mname achieves the best performance on over 90\% of locations for both tasks. For respiratory diseases, \mname has 12.4\% improved MAPE on average than the best baseline. For tumors prediction, \mname achieves the best performance on 1645 locations, and the average MAPE improvement is 10.7\%. Even for the locations that  baselines perform the best, the MAPE gap is small.
\\

\subsection*{Q3. Performance with Iterative Training}
\begin{figure}[h!]
    \centering
    \includegraphics[width=0.95\columnwidth]{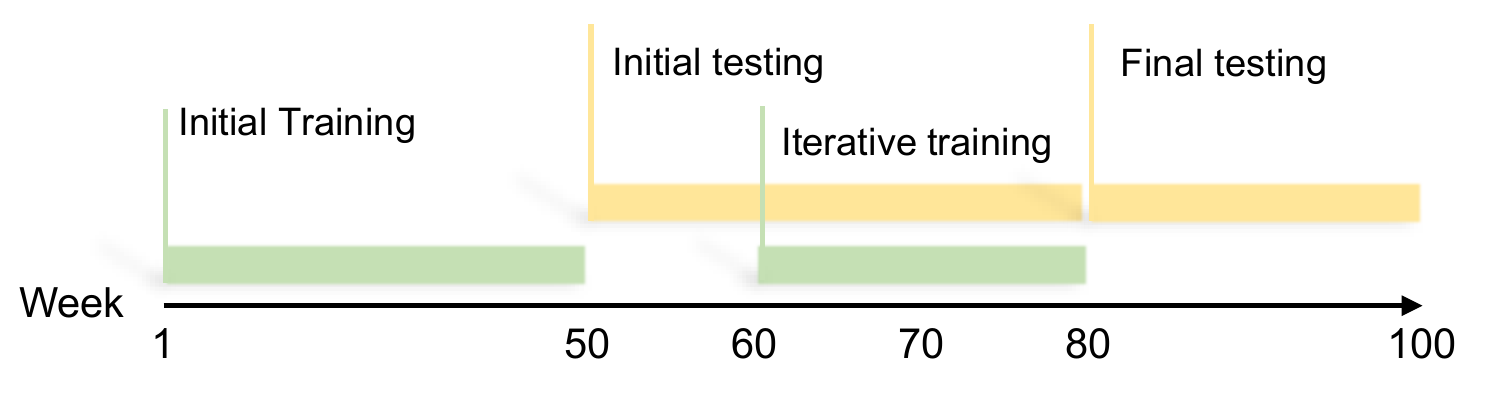}
    \caption{Data splitting process to simulate iterative training}
    \label{fig:time_split}
\end{figure}

To evaluate whether the iterative training of \mname improves the efficiency of model updating, we  simulate the following deployment setting:  train on the original dataset (week 1-50), and deploy into practice (week 50-80). Then refresh the model using newly collected data during the deployment phase (week 60-80). Finally, we re-deploy the model to test the performance (week 80-100). The entire splitting process is shown in Fig.~\ref{fig:time_split}.

We report the  performance on the final testing phase  in Table~\ref{res_iter}.  \mname  outperforms baselines, achieving 39\% lower RMSE, 34\% lower MAE, and 25.6\% lower MAPE on respiratory disease prediction. and 70\% lower RMSE and 49\% lower MAE on tumor prediction, compared with the best baseline. Compared with the reduced model \mname-$\mathcal{L}_\alpha$, we can see the alignment loss can indeed help improve the predictive performance by providing RNN with a better initialization.

\begin{table}[h!]
    \caption{Prediction performance with iterative traing}
\begin{tabular}{lcccc}
\toprule
\multicolumn{5}{c}{\textbf{Respiratory diseases}} \\
                         \textbf{Model}   & \textbf{RMSE ($\times10^5$)} & \textbf{MAE} & \textbf{MAPE} & \textbf{P-value} \\ \hline
 GRU             & 14.77 & 495.4 & 52.5 & 0.0\\
                          GMAN             & 14.17 & 450.5 & 47.3 & 0.0\\
                          ASTGCN             & 12.45 & 432.7 & 46.2 & 3e-20\\
                    EvolveGCN             & 12.18 & 429.7 & 43.1 & 1e-18\\
                          STAN             & 13.90 & 448.5 & 47.0 & 0.0\\
                
                           ColaGNN        & 11.82 & 416.1 & 39.8  &9e-17\\ \rowcolor{Gray}
\mname-$\mathcal{L_\alpha}$&11.43        &    355.3         &    33.5 & 5e-10\\ \rowcolor{Gray}
 \mname&\textbf{7.23}          &    \textbf{275.6}           &    \textbf{29.6} & -\\
\midrule
\multicolumn{5}{c}{\textbf{Tumors}} \\
                          \textbf{Model}   & \textbf{RMSE ($\times10^5$)} & \textbf{MAE} & \textbf{MAPE} & \textbf{P-value} \\ \hline
 GRU             & 18.95 & 397.6 & 52.9 & 8e-23\\
                          GMAN             & 18.99 & 401.6 & 53.2 & 0.0\\
                         ASTGCN             & 19.72 & 401.3 & 55.6 & 0.0\\
        EvolveGCN             & 17.26 & 385.1 & 50.0 & 0.0\\
                          STAN             & 10.21 & 304.7 & 40.8 & 6e-15\\
                          ColaGNN        & 8.75 & 264.4 & 35.3  & 1e-8\\ \rowcolor{Gray}
\mname-$\mathcal{L_\alpha}$&4.92        &    182.7          &    35.8 & 3e-4\\
 \rowcolor{Gray}
 \mname&\textbf{2.60}          &    \textbf{135.4}           &    \textbf{34.7} & -\\
\bottomrule
\end{tabular}
\label{res_iter}
\end{table}

\begin{figure}[h!]
    \centering
    \includegraphics[width=0.95\columnwidth]{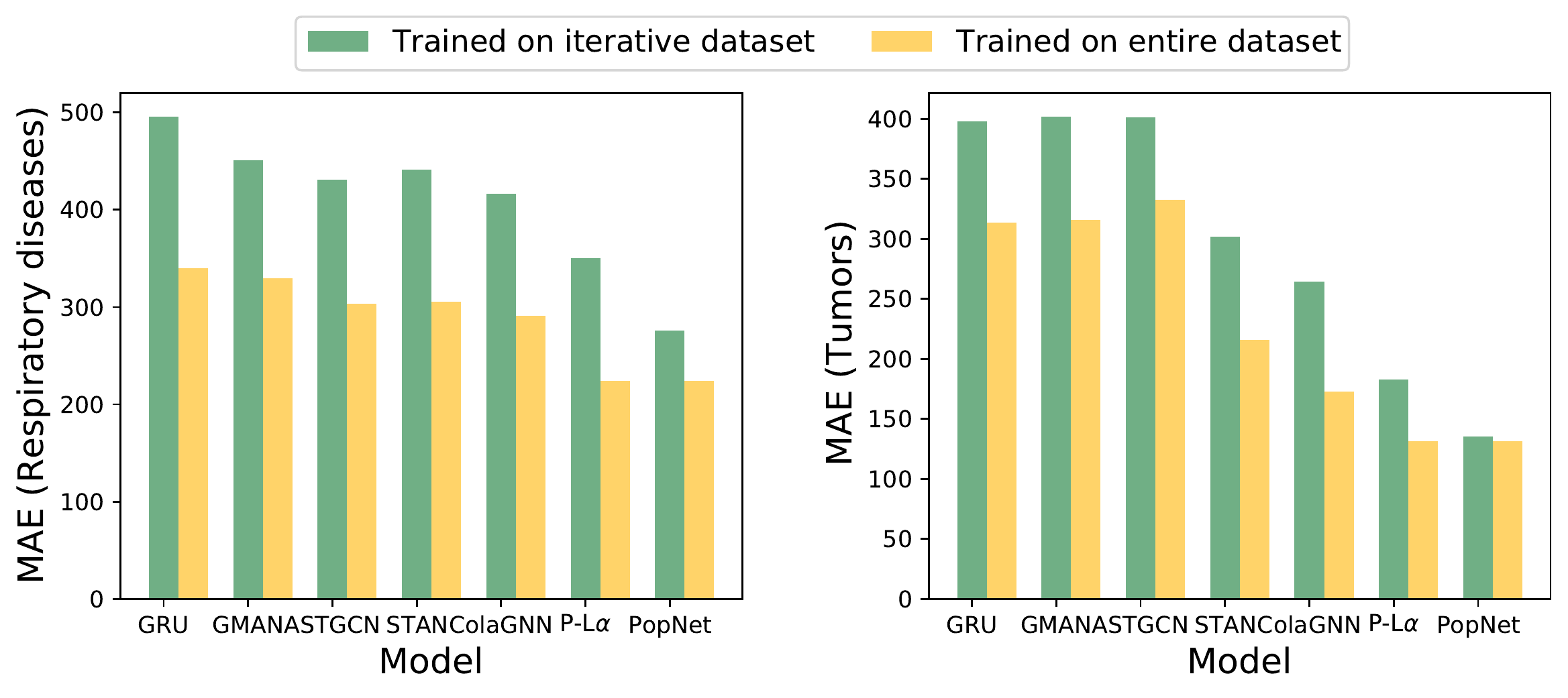}
    \caption{Prediction MAE for models trained on entire dataset and iterative dataset}
    \label{fig:iter}
    \vskip -2em
\end{figure}

To further evaluate how $\mathcal{L}_\alpha$ can help improve the iterative training, in Fig.~\ref{fig:iter}, we compare the iterative training results with the models that are trained on the entire sequence (results in Table \ref{tumor_res}). Compared with the results reported in Table~\ref{res_iter} (yellow bars), the results in Table \ref{tumor_res} (green bars) are obtained using the same test set but trained on the entire historical sequence (week 1-80). The P-$\mathcal{L}_\alpha$ indicates the reduced model \mname-$\mathcal{L_\alpha}$.

The figure shows that the models trained on the entire dataset can achieve lower MAE because the model can access the entire historical data, which makes the models can extract and utilize more historical patterns. However, compared to all baselines and the reduced model, the performance gap of \mname is much smaller. Compared to the model trained under the iterative training setting, \mname trained on entire sequences achieves 23\% higher MAE on respiratory diseases prediction and only 3\% higher MAE on tumors prediction. In comparison, the best baseline model trained on entire sequences achieves 41\% higher MAE on respiratory diseases prediction and 53\% higher MAE on tumors prediction. This indicates that by aligning the TIE and hidden states of RNN, the model can indeed utilize historical patterns without accessing the original sequences. 

Since the entire sequence is four times longer than the iterative training data, training on the entire dataset could be costly for time and memory. For example, for tumor disease, training a regular spatial-temporal model on the entire dataset generally requires more than 12 GB memory and the average training time is about 5 seconds per epoch. But it only requires less than 4 GB memory and 2 seconds per epoch to train the same model on the iterative dataset. \mname can achieve almost equivalent performance using just iterative training data, which can be useful for real-world applications and efficient for long-sequence data.

\subsection*{Q4. Performance with Longer Prediction Window}
Long-term prediction is also significant for disease prediction in practice. In this work, we also explore the capability of \mname for long-term disease prediction. We change the output size to make \mname and other baseline models predict future 5 weeks. We report the performance and the p-value of MAE in Table~\ref{res_long}.

\begin{table}[h!]
\vskip -1em
    \caption{Long-term prediction (window $=$ 5 weeks)}
\begin{tabular}{lcccc}
\toprule
\multicolumn{5}{c}{\textbf{Respiratory Diseases}} \\
                         \textbf{Model}   & \textbf{RMSE ($\times10^5$)} & \textbf{MAE} & \textbf{MAPE} & \textbf{P-value} \\ \hline
 GRU             & 16.90 & 491.4 & 34.7 & 7e-21\\
                          GMAN             & 17.76 & 440.3 & 35.2 & 6e-23\\
                          ASTGCN             & 15.23 & 410.3 & 33.4 & 2e-15\\
                          STAN             & 9.30 & 380.5 & 31.5 & 4e-8\\
                           ColaGNN        & 9.78 & 384.1 & 31.7  &8e-8\\ \rowcolor{Gray}
 \mname&\textbf{5.71}          &    \textbf{255.4}           &    \textbf{27.8} & -\\
\multicolumn{5}{c}{\textbf{Tumors}} \\
                          \textbf{Model}   & \textbf{RMSE ($\times10^5$)} & \textbf{MAE} & \textbf{MAPE} & \textbf{P-value} \\ \hline
 GRU             & 21.56 & 452.6 & 38.8 & 0.0\\
                          GMAN             & 19.35 & 420.1 & 35.9 & 5e-20\\
                         ASTGCN             & 19.78 & 425.1 & 36.0 & 6e-23\\
                          STAN             & 6.72 & 240.3 & 34.1 & 3e-9\\
                          ColaGNN        & 5.37 & 231.9 & 33.5  & 5e-7\\ \rowcolor{Gray}
 \mname&\textbf{2.79}          &    \textbf{138.4}           &    \textbf{31.5} & -\\
\bottomrule
\end{tabular}
\label{res_long}
\vskip -1em
\end{table}

The results show that the MAE rises as the prediction window increases since it becomes more difficult to predict longer future trends. However, \mname still has the lowest MAE increase ratio.

\begin{figure}[h!]
    \centering
    \includegraphics[width=0.95\columnwidth]{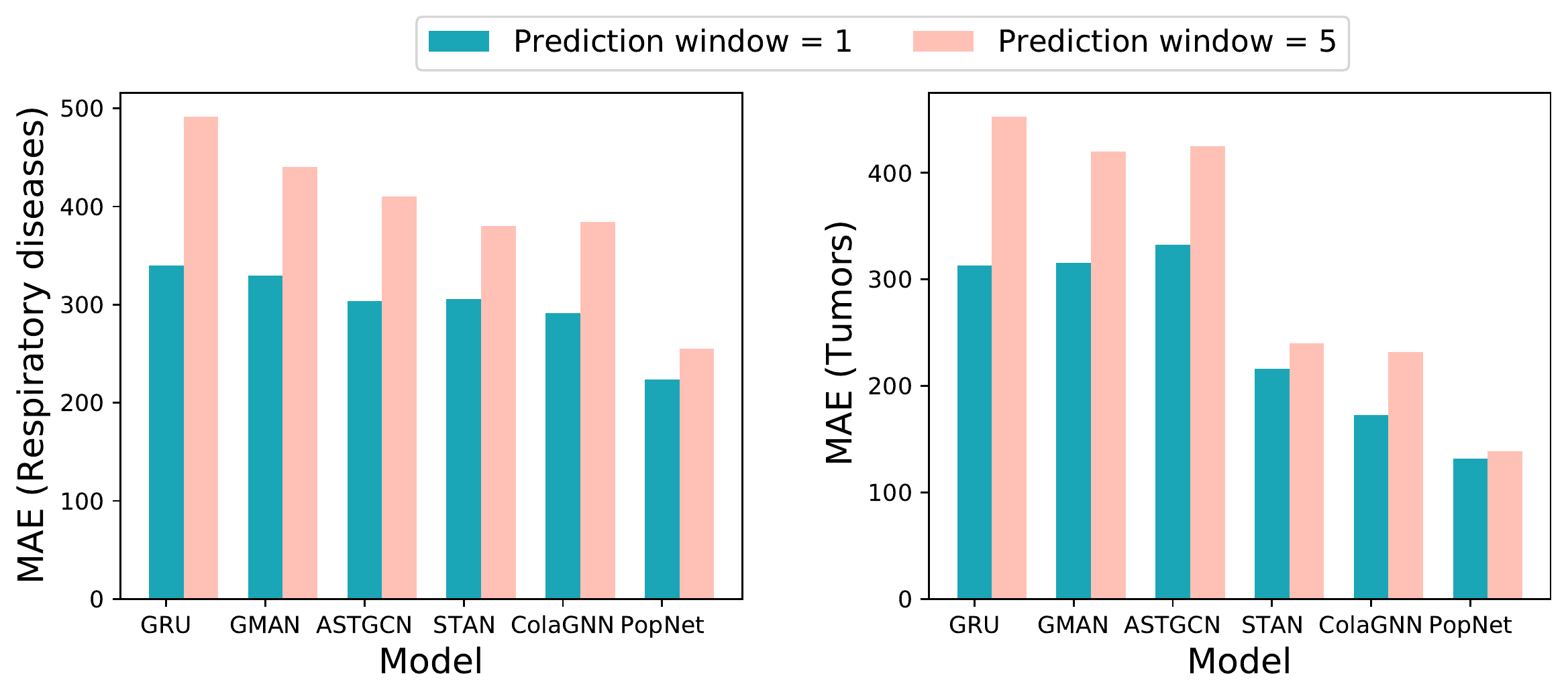}
    \caption{Prediction MAE under prediction window 1 and 5}
    \label{fig:gap}
    \vskip -2em
\end{figure}

For respiratory diseases, the MAE of \mname increases 14\% as the prediction window length increases from 1 to 5, while the baseline model STAN increases 24\% and ColaGNN increases 31\%. For tumors, the MAE of STAN and ASTGCN increase 11\% and 24\%, respectively, while \mname only increases 6\%. The results show that \mname can consistently outperform all other baseline models under different lengths of prediction window. A longer prediction window has less effect on the predictive performance of \mname. To illustrate more clearly, we draw the performance gap of all baseline models with different prediction windows in Fig.~\ref{fig:gap}. The figure shows that compared to baseline models, \mname can achieve a significantly smaller error gap as the prediction window increases on two diseases. This indicates that \mname is also suitable for long-term prediction tasks. 

\section{Conclusion}
In this work, we propose  \mname for real-time population-level disease prediction with considering data latency. \mname uses two separate systems to model real-time and updated disease statistics data, and then adaptively fuses the two systems using both spatial and  temporal latency-aware cross-graph attention. We augment the latency-aware attention with spatial and temporal information embeddings to adaptively extract and utilize geographical and temporal progression features. We also conducted extensive experiments across multiple real-world claims datasets. \mname outperforms leading spatial-temporal models in all metrics and shows the 
promising utility and efficacy in population-level disease prediction. In future works, we will use more flexible way to generate better location graph instead of using hard defined edge weights, which is the major limitation of this work.

\section{Acknowledgments}
This work was supported by IQVIA, NSF award SCH-2014438, PPoSS 2028839, IIS-1838042, NIH award R01 1R01NS107291-01 and OSF Healthcare.

\newpage
\bibliographystyle{format}
\bibliography{references}

\newpage
\appendix
\section{Dataset details}

\subsection{Real-world dataset}
In this section, we report the basic statistics and disease categories in the real-world claims dataset. The patients' claims data are collected from 2952 counties in the US starting from 2018. We aggregate the codes into 17 disease categories (A00-Q99) and 4 other categories (R00-Z99) according to the ICD-10 coding. The detailed disease category is reported in Table~\ref{tab:category}.

\begin{table}[h!]
\resizebox{1\columnwidth}{!}{
    \begin{tabular}{lll}
    \toprule
    ICD code & Category description & Avg. cases\\
    \midrule
    A00-B99 & Certain infectious and parasitic diseases & 314.9 \\
    B00-D49 & Tumors & 966.1 \\
    D50-D89 & \tabincell{l}{Diseases of the blood, blood-forming \\organs and immune mechanism} & 344.8 \\
    E00-E89 & Endocrine, nutritional and metabolic diseases & 1508.8 \\
    F01-F99 & \tabincell{l}{Mental, Behavioral and mental disorders} & 1470.1 \\
    G00-G99 & Diseases of the nervous system & 744.2 \\
    H00-H59 & Diseases of the eye and adnexa & 385.8 \\
    H60-H95 & Diseases of the ear and mastoid process & 192.2 \\
    I00-I99 & Diseases of the circulatory system & 1717.2 \\
    J00-J99 & Diseases of the respiratory system & 1774.7 \\
    K00-K95 & Diseases of the digestive system & 653.1\\
    L00-L99 & Diseases of the skin and subcutaneous tissue & 496.9 \\
    M00-M99 & \tabincell{l}{Diseases of the musculoskeletal system \\and connective tissue} & 2226.7 \\
    N00-N99 & Diseases of the genitourinary system & 897.0 \\
    O00-O99 & Pregnancy, childbirth and the puerperium & 152.6 \\
    P00-P96 & \tabincell{l}{Certain conditions originating \\ in the perinatal period} & 45.9 \\
    Q00-Q99 & \tabincell{l}{Congenital malformations, deformations \\and chromosomal abnormalities} & 80.2 \\
    R00-R99 & \tabincell{l}{Symptoms, signs and abnormal clinical \\and laboratory findings} & 2480.2 \\
    S00-T88 & \tabincell{l}{Injury, poisoning and certain other \\consequences of external causes} & 683.3 \\
    V00-Y99 & External causes of morbidity & 12.4 \\
    Z00-Z99 & \tabincell{l}{Factors influencing health status \\and contact with health services} & 2653.2 \\
    \bottomrule
    \end{tabular}}
    \caption{ICD codes and disease category descriptions}
    \label{tab:category}
\end{table}

Due to space limitations, we only report the detailed data statistics of two diseases (i.e., tumors and respiratory diseases) reported in the main text. The statistics are shown in Table~\ref{tab:stat}. The spatial features include populations, number of hospitals, number of ICU beds, longitude, latitude and annual income.

\begin{table}[h!]
    \begin{tabular}{lc}
    \toprule
    Respiratory diseases& \\
    \midrule
        \# of locations & 1693 \\
         \# of features & 21 \\
         \# of edges & 4521 \\
         Avg. \# of edges per nodes & 2.67 \\
         \# of sequence length & 63 \\
         Avg. \# of target cases & 1774.7 \\
         Avg. \# of update frequencies & 9.3\\
    \toprule
    Tumors &\\
    \midrule
    \# of locations & 1829 \\
         \# of features & 22 \\
         \# of edges & 4884 \\
         Avg. \# of edges per nodes & 2.67 \\
         \# of sequence length & 63 \\
         Avg. \# of target cases & 966.1 \\
         Avg. \# of update frequencies & 3.3 \\
    \bottomrule
    \end{tabular}
    \caption{Data statistics for respiratory disease and tumors dataset}
    \label{tab:stat}
\end{table}

\subsection{Synthetic dataset}
We construct a synthetic dataset from a real-world disease dataset. We first randomly aggregate the data in the real-world dataset from different locations to generate data sequences for the real-time data and prediction targets. For each location, We randomly aggregate data from 1-5 neighboring locations. Then for each timestep, we use up-sampling and down-sampling to aggregate data from 1-3 continuous timesteps while keeping the length of data does not change. Then we add random Gaussian noise ($\mu=0, \sigma=1$) to the aggregated data. For the updated data, we assume all locations are updated at regular intervals for the sake of simplicity. We use the same strategy as the real-time data to generate the updated data. The basic statistics of the synthetic dataset are shown in Table~\ref{tab:syn_stat}.

\begin{table}[h!]
    \begin{tabular}{lc}
    \toprule
        \# of locations & 1015 \\
         \# of features & 22 \\
         \# of edges & 6410 \\
         Avg. \# of edges per nodes & 6.32 \\
         \# of sequence length & 63 \\
         Avg. \# of target cases & 1098.7 \\
         Avg. \# of update frequencies & 5.4 \\
    \bottomrule
    \end{tabular}
    \caption{Synthetic data statistics}
    \label{tab:syn_stat}
\end{table}

\section{Prediction performance for all disease categories}

In this section, we report the predictive performance of \mname for all disease categories. Due to space limitations, we only select two baseline models (i.e., STAN and ColaGNN) to compare, which have generally better performance. For some disease categories such as \textit{Certain conditions originating in the perinatal period}, some locations have no case at most timesteps, so these disease datasets have fewer locations to predict. We report the test MAE in Table~\ref{tab:all_performance}. \mname can outperform two baselines for all disease code categories.
\begin{table}[h!]
    \begin{tabular}{lcccc}
    \toprule
    ICD code & \# of locations & STAN & ColaGNN & \mname \\
    \midrule
    A00-B99 &1,654 & 173.6 & 152.7 & \textbf{53.7}\\
    D50-D89 &1,545 & 210.5 & 132.2 & \textbf{70.1}\\
    E00-E89 &1,793 & 310.0 & 285.4 & \textbf{268.4}\\
    F01-F99 &1,675 & 325.4 & 301.3 & \textbf{287.4}\\
    G00-G99 &1,733 & 225.9 & 269.1 & \textbf{124.1}\\
    H00-H59 &1,530 & 129.3 & 150.7 & \textbf{113.4}\\
    H60-H95 &1,551 & 97.1 & 85.2 & \textbf{43.9}\\
    I00-I99 &1,642 & 439.5 & 405.5 & \textbf{321.3}\\
    K00-K95 &1,557 & 280.7 & 350.6 & \textbf{124.8}\\
    L00-L99 &1,424 & 192.9 & 245.8 & \textbf{122.9}\\
    M00-M99 &1,789 & 371.2 & 369.7 & \textbf{355.5}\\
    N00-N99 &1,673 & 288.9 & 252.0 & \textbf{163.8}\\
    O00-O99 &1,512 & 170.2 & 95.3 & \textbf{31.4 }\\
    P00-P96 &978 & 115.8 & 81.9 & \textbf{19.8}\\
    Q00-Q99 &1,340 & 157.4 & 109.7 & \textbf{22.7}\\
    \bottomrule
    \end{tabular}
    \caption{Test MAE for all disease categories}
    \label{tab:all_performance}
\end{table}

\section{Performance on Synthetic Dataset}

\begin{table}[h!]
    \caption{Prediction performance on synthetic dataset}
\begin{tabular}{lcccc}
\toprule
                          \textbf{Model}   & \textbf{RMSE ($\times10^5$)} & \textbf{MAE} & \textbf{MAPE} & \textbf{P-value}\\ \hline
SARIMAX             & 6.42 & 252.4 & 60.0 & 0.0\\
 GRU             & 2.37 & 130.3 & 36.7 & 1e-3\\
                          GMAN             & 3.52 & 142.4 & 41.3 & 5e-4\\
                          ASTGCN             & 2.56 & 128.4 & 37.2 & 8e-4 \\
                          STAN             & 2.33 & 122.5 & 35.7 & 2e-3 \\
                          ColaGNN  & 2.21 & 115.9 & 34.2 & 3e-3  \\ 
\rowcolor{Gray} \mname-LAtt       &   2.55   &  135.2  &  37.2 & 9e-3  \\
\rowcolor{Gray} \mname-TLAtt        &    2.08   & 106.9 &  33.8 & 1e-3  \\ 
\rowcolor{Gray} \mname-SLAtt        &    2.15   & 113.8 &  34.0 & 2e-3  \\ 
\rowcolor{Gray} \mname&\textbf{1.78}          &    \textbf{97.9}           &    \textbf{33.0} & - \\
\bottomrule
\end{tabular}
\label{syn_res}
\end{table}

We also conducted  experiments  on the artificially generated synthetic dataset, and report results in Table~\ref{syn_res}. From the results, \mname outperforms all  baselines with a $p=0.001$ significance level. Compared with the best baseline  ColaGNN, \mname has 19.5\% lower RMSE, 15.5\% lower MAE, and 4\% lower MAPE. The SARIMAX model does not perform well on the synthetic dataset since autoregression models are difficult to fit random noises in the data.

\section{Implementation details}
All methods are implemented in PyTorch~\cite{paszke2019pytorch} and trained on an Ubuntu 16.04 with 64GB memory and a Tesla V100 GPU. We use Adam optimizer \cite{kingma2014adam} with a learning rate of 0.001 and trained for 200 epochs. 

For hyper-parameter settings of each baseline model, our principle is as follows: For some hyper-parameter, we will use the recommended setting if available in the original paper. Otherwise, we determine its value by grid search on the validation set.
 
\begin{itemize}
    \item \textbf{SARIMAX} stands for seasonal autoregressive integrated moving average, which is a popular time series prediction model.  SARIMAX considers seasonal influence with exogenous variables, making it more suitable for our disease prediction task. We use grid-search to determine the hyperparameters of the model at each location.
    \item \textbf{GRU}. We use GRU to conduct temporal prediction without considering the spatial relationships. GRU model cannot utilize spatial relationships and locations are regarded as independent samples to train the GRU model. The hidden units of the GRU cell are set to 128.
    \item \textbf{GMAN} is a recently published spatial-temporal prediction model for traffic prediction. It uses an encoder-decoder structure with spatial-temporal attention to predict future traffic status. The number of attention blocks is set to 3, the dimensionality of each attention head is set to 64, and the number of attention head is 4.
    \item \textbf{ASTGCN} is a recently published spatial-temporal prediction model for traffic prediction. It applies additional convolutional layers and attention mechanisms on GCN. The number of convolutional kernels is set to 64, and the kernel size is set to 3.
    \item \textbf{EvolveGCN} is a general spatial-temporal prediction model. It adapts the graph convolutional network (GCN) model along the temporal dimension without resorting to node embeddings and uses an RNN to evolve the GCN parameters.  The hidden units of the GRU cell are set to 128, and the dimensionality of GNN is set to 64.
    \item \textbf{STAN} is a hybrid deep learning and epidemiology spatial-temporal model for epidemic and pandemic prediction. STAN also constructs a location graph based on geographic similarity and uses graph attention network and RNN to predict future cases. Since we do not constraint our prediction target is an infectious disease, we remove the disease transmission dynamics regularization in STAN. The dimensionality of GAT is set to 64 for respiratory diseases prediction and tumors prediction, 128 for the synthetic dataset. The number of hidden units of GRU cell is set to 128.
    \item \textbf{ColaGNN} is a spatial-temporal pandemics prediction model, which uses a location graph to extract spatial relationships for predicting pandemics. The number of hidden units of GRU cell is set to 128. The number of convolutional kernels is set to 64, and the kernel size is set to 3.
    \item \textbf{\mname}. The $\alpha, \beta, \gamma$ is set to 0.35, 0.37, 30. We set the kernel size of convolutional layers to 16 and kernel size to 3. We use a set of dilation rate $\phi= [1,3,5]$. The dimensionality of the GAT layer and attention head is set to 32. We use 2 attention heads for respiratory diseases prediction and tumors prediction, 1 for the synthetic dataset. The hidden units of GRU are set to 256. The dimensionality of MLP is set to 128.
\end{itemize}

We also use a dropout layer~\cite{srivastava2014dropout} before the output layer to prevent overfitting. The dropout rate is set to 0.5.

\end{document}